\documentclass[aps,prd,twocolumn,amsfonts]{revtex4}
\usepackage{amsmath}
\usepackage{graphicx}
\def\D{{\cal D}}
\def\R{{\cal R}}
\def\G{{\cal G}}
\def\H{{\cal H}}
\def\g{\,^{(5)\!}g}
\def\m{{(\!m\!)}}
\def\vu{^{(5)\!}}
\newcommand{\be}{\begin{equation}}
\newcommand{\ee}{\end{equation}}
\newcommand{\bea}{\begin{eqnarray}}
\newcommand{\eea}{\end{eqnarray}}
\begin{document}

\title{Cosmological perturbations from braneworld inflation
with a Gauss-Bonnet term}

\author{Jean-Francois Dufaux$^1$, James E. Lidsey$^2$,
Roy Maartens$^3$, M Sami$^4$}

\affiliation{\vspace*{0.2cm}$^1$LPT, Universit\'e de Paris-Sud,
91405 Orsay, France\\ $^2$Astronomy Unit, School of Mathematical
Sciences, \\ Queen Mary, University of London, London~E1~4NS, UK\\
$^3$Institute of Cosmology \& Gravitation, University of
Portsmouth, Portsmouth~PO1~2EG, UK
\\ $^4$Inter-University Centre for Astronomy \& Astrophysics,
Pune, India \vspace*{0.2cm}}

\date{\today}

\begin{abstract}

Braneworld inflation is a phenomenology related to string theory that
describes high-energy modifications to general relativistic
inflation. The observable universe is a braneworld embedded in
5-dimensional anti de Sitter spacetime. When the 5-dimensional
action is Einstein-Hilbert, we have a Randall-Sundrum type
braneworld. The amplitude of tensor and scalar perturbations from
inflation is strongly increased relative to the standard results,
although the ratio of tensor to scalar amplitudes still obeys the
standard consistency relation. If a Gauss-Bonnet term is included
in the action, as a high-energy correction motivated by string
theory, we show that there are important changes to the
Randall-Sundrum case. We give an exact analysis of the tensor
perturbations. They satisfy the same wave equation and have the
same spectrum as in the Randall-Sundrum case, but the Gauss-Bonnet
change to the junction conditions leads to a modified amplitude of
gravitational waves. The amplitude is no longer monotonically
increasing with energy scale, but decreases asymptotically after
an initial rise above the standard level. Using an approximation
that neglects bulk effects, we show that the amplitude of scalar
perturbations has a qualitatively similar behaviour to the tensor
amplitude. In addition, the tensor to scalar ratio breaks the
standard consistency relation.

\end{abstract}


\maketitle

\section{Introduction}

In recent years, there has been considerable interest in the
possibility that our observable four-dimensional (4D) universe may
be viewed as a brane hypersurface embedded in a higher-dimensional
bulk space. Of particular importance is the Randall-Sundrum (RS)
model, where a single, positive-tension brane is embedded in a
five-dimensional (5D) anti de Sitter $({\rm AdS}_5)$
spacetime~\cite{RSII}. (For recent reviews, see
Ref.~\cite{royreview}.) Although the fifth dimension may be
infinite in extent, the zero-mode of the 5D graviton,
corresponding to 4D gravitational waves, is localized at low
energies on the brane due to the warped geometry of the bulk. This
property can also be understood within the context of the AdS/CFT
correspondence~\cite{adscft}, where the RS model is viewed as 4D
gravity coupled to a conformal field theory (CFT)~\cite{hrr}.

A natural extension of the RS model that is motivated by
string/M~theory considerations, is to include higher-order
curvature invariants in the bulk action. Such terms arise in the
AdS/CFT correspondence as next-to-leading order corrections to the
CFT~\cite{largeN}. The Gauss-Bonnet (GB) combination of curvature
invariants is of particular relevance in five dimensions, since it
represents the unique combination that leads to second-order
gravitational field equations in the bulk metric and since the
field equations contain only linear second derivatives~\cite{d86}.
A GB term may also arise as the next-to-leading order correction
in the heterotic string effective action, and it is ghost-free
about flat spacetime~\cite{stringGB}. Moreover, the graviton zero
mode remains localized in the GB braneworld~\cite{onlylocal} and
deviations from Newton's law at low energies are less pronounced
than in the RS model~\cite{milder}.

From an observational perspective, there is now strong evidence
that the very early universe underwent an epoch of accelerated
(inflationary) expansion~\cite{wmap}. During inflation, light
fields such as the graviton are quantum-mechanically excited and
acquire nearly scale-invariant fluctuations~\cite{gw}. The
resulting spectrum of primordial gravitational waves could be
detectable from its imprint on the polarization of the cosmic
microwave background (CMB)~\cite{Bmode}. Such a detection would
open a unique window into the physics of the very early universe.
The evolution of gravitational waves during slow-roll inflation
has been determined in the RS scenario~\cite{lmw}. At high
energies, the amplitude is enhanced relative to the standard
result in 4D Einstein gravity. In view of the above developments,
therefore, it is important to determine the properties of tensor
perturbations generated during inflation in the GB braneworld. We
show that significant changes to the RS case are introduced by the
GB term, even when the GB corrections are very small relative to
the Einstein-Hilbert terms.

\section{Field Equations}

For a 5D bulk with Einstein-Gauss-Bonnet gravity, containing a 4D
brane, the gravitational action is
 \bea
{\cal S} &=& \frac{1}{2\kappa_5^2} \int d^5x \sqrt{-\,\vu g}
\left[-2\Lambda_5+ {\cal R} \right. \nonumber\\
&&\left.~{}+\alpha\, \left({\cal R}^2-4 {\cal R}_{ab}{\cal
R}^{ab}+ {\cal R}_{abcd}{\cal R}^{abcd}\right) \right] \nonumber\\
&&~{} - \int_{\rm brane} d^4x\, \sqrt{-g}\, \sigma\,,
\label{action}
 \eea
where $x^a=(x^\mu,z)$, $g_{ab}=\,\vu g_{ab}-n_an_b$ is the induced
metric, with $n^a$ the unit normal to the brane, $\sigma\, (>0)$
is the brane tension, and $\Lambda_5\,(<0)$ is the bulk
cosmological constant. The fundamental energy scale of gravity is
the 5D scale $M_5$, where $\kappa_5^2=8\pi/M_5^3$. The Planck
scale $M_4\sim 10^{16}~$TeV is an effective scale, describing
gravity on the brane at low energies, and typically $M_4 \gg M_5$.

The GB term may be thought of as the lowest-order stringy
correction to the 5D Einstein-Hilbert action, with coupling
constant $\alpha>0$. In this case, $\alpha|\R^2|\ll|\R|$, so that
 \be
\alpha \ll \ell^2\,, \label{al}
 \ee
where $\ell$ is the bulk curvature scale, $|\R|\sim \ell^{-2}$.
The RS type models are recovered for $\alpha=0$.

The 5D field equations following from the bulk action are
\begin{eqnarray}
\G_{ab} &=& -\Lambda_5 \g_{ab}+{\alpha\over 2}{\H}_{ab}\,,\label{fe}\\
\H_{ab} &=& \left[\R^2-4\R_{cd}\R^{cd}+\R_{cdef}\R^{cdef}
\right]\g_{ab} \nonumber\\&&~{}-4 \left[ \R\R_{ab}-
2\R_{ac}\R^c{}_b\right.
\nonumber\\&&~\left.{}-2\R_{acbd}\R^{cd}+\R_{acde}\R_b{}^{cde}
\right]\,.\label{5d}
\end{eqnarray}
The junction conditions at the brane, assuming mirror ($Z_2$)
symmetry, are~\cite{junc}
\begin{eqnarray}
K_{\mu\nu}-Kg_{\mu\nu} &=&
-{\kappa_5^2\over2}\left(T_{\mu\nu}-\sigma g_{\mu\nu} \right)
\nonumber\\&&{} -2\alpha\left(\! Q_{\mu\nu}
-{1\over3}Qg_{\mu\nu}\! \right)\, ,\label{gjc}
\end{eqnarray}
where
\begin{eqnarray}
Q_{\mu\nu} &=& 2KK_{\mu\alpha}K^\alpha{}_\nu -2
K_{\mu\alpha}K^{\alpha\beta}K_{\beta\nu} \nonumber\\&&~{} + \left(
K_{\alpha\beta}K^{\alpha\beta}-K^2 \right)K_{\mu\nu}+ 2KR_{\mu\nu}
\nonumber\\&&~{} +RK_{\mu\nu}+2K^{\alpha\beta}
R_{\mu\alpha\beta\nu} -4R_{\mu\alpha}K^\alpha{}_\nu\,.
\end{eqnarray}
Here the curvature tenors are those of the 4D induced metric
$g_{\mu\nu}$, $K_{\mu\nu}$ is the extrinsic curvature and
$T_{\mu\nu}$ is the brane energy-momentum tensor. For a vacuum
bulk, the conservation equations hold:
\begin{equation}
\label{tmunu} \nabla^\nu T_{\mu\nu}=0\,.
\end{equation}

An AdS$_5$ bulk satisfies the 5D field equations, with
\begin{eqnarray}
\bar{\R}_{abcd}&=& -{1\over \ell^2}\left[ \,^{(5)\!}\bar{g}_{ac}
\,^{(5)\!}\bar{g}_{bd} - \,^{(5)\!}\bar{g}_{ad}
\,^{(5)\!}\bar{g}_{bc} \right],\\ \bar{\G}_{ab} &=& {6\over
\ell^2} \,^{(5)\!}\bar{g}_{ab}=-\Lambda_5 \,^{(5)\!}\bar{g}_{ab}
+{\alpha\over 2}\bar{\H}_{ab}\,,\label{bg}\\ \bar{\H}_{ab} &=& {24
\over \ell^4}\,^{(5)\!}\bar{g}_{ab}\,.
\end{eqnarray}
It follows that
\begin{eqnarray}
\Lambda_5 &=& -{6\over \ell^2} +{12\alpha\over
\ell^4}\,,\label{lam}
\\
{1\over \ell^2} &\equiv& \mu^2={1\over 4\alpha} \left[1 - \sqrt{
1+{4\over3} \alpha\Lambda_5} \right] \,,\label{ell}
\end{eqnarray}
where we choose in Eq.~(\ref{ell}) the branch with an RS limit,
and $\mu$ is the energy scale associated with $\ell$. This reduces
to the RS relation $1/\ell^2=-\Lambda_5/6$ when $\alpha=0$. Note
that there is an upper limit to the GB coupling from
Eq.~(\ref{ell}):
\begin{equation}\label{lim}
{\alpha} < {\ell^2 \over 4}\,,
\end{equation}
which in particular ensures that $\Lambda_5<0$.

A Friedman-Robertson-Walker (FRW) brane in an AdS$_5$ bulk is a
solution to the field and junction equations~\cite{cd1}. The
modified Friedman equation on the (spatially flat) brane
is~\cite{cd1,mt}
 \be
\kappa_5^2(\rho+\sigma) = 2\sqrt{H^2+\mu^2}\left[3-4\alpha\mu^2
+8\alpha H^2\right]. \label{mf}
 \ee
This may be rewritten in the useful form~\cite{ln}
 \bea
H^2 &=& {1\over
4\alpha}\left[(1-4\alpha\mu^2)\cosh\left({2\chi\over3}
\right)-1\right]\,,\label{mfe}\\
\label{chi} \kappa_5^2(\rho+\sigma) &=&
\left[{{2(1-4\alpha\mu^2)^3} \over {\alpha} }\right]^{1/2}
\sinh\chi\,,
 \eea
where $\chi$ is a dimensionless measure of the energy density.
Note that the limit in Eq.~(\ref{lim}) is necessary for $H^2$ to
be non-negative.

When $\rho=0=H$ in Eq.~(\ref{mf}) we recover the expression for
the critical brane tension which achieves zero cosmological
constant on the brane,
 \bea
\kappa_5^2\sigma &=& 2\mu(3-4\alpha\mu^2)\,. \label{sig'}
 \eea
The effective 4D Newton constant is given by~\cite{onlylocal}
\begin{equation}\label{k4k5}
{\kappa_4^2  }= {\mu \over (1+4\alpha\mu^2) }\,\kappa_5^2\,.
\end{equation}
When Eq.~(\ref{al}) holds, this implies $M_5^3\approx M_4^2/\ell$.
Table-top experiments to probe deviations from Newton's law
currently imply $\ell\lesssim 0.1~$mm, so that $M_5 \gtrsim
10^5~$TeV, and $\sigma \gtrsim (1~{\rm TeV})^4$, by
Eqs.~(\ref{al}) and (\ref{sig'}).

The modified Friedman equation~(\ref{mfe}), together with
Eq.~(\ref{chi}), shows that there is a characteristic GB energy
scale,
 \be \label{m*}
m_\alpha= \left[{{2(1-4\alpha\mu^2)^3} \over {\alpha} \kappa_5^4
}\right]^{1/8}\,,
 \ee
such that the GB high energy regime ($\chi\gg1$) is $\rho+\sigma
\gg m_\alpha^4$. If we consider the GB term in the action as a
correction to RS gravity, then $m_\alpha$ is greater than the RS
energy scale $m_\sigma=\sigma^{1/4}$, which marks the transition
to RS high-energy corrections to 4D general relativity. By
Eq.~(\ref{sig'}), this requires $3\beta^3-12\beta^2+15\beta-2<0$
where $\beta \equiv 4\alpha\mu^2$. Thus (to 2 significant
figures),
 \be \label{lim2}
m_\sigma< m_\alpha~\Rightarrow~\alpha\mu^2 < 0.038\,,
 \ee
which is consistent with Eq.~(\ref{al}).

Expanding Eq.~(\ref{mfe}) in $\chi$, we find three regimes for the
dynamical history of the brane universe:\\ the GB regime,
 \be
\rho\gg m_\alpha^4~ \Rightarrow ~ H^2\approx \left[ {\kappa_5^2
\over 16\alpha}\, \rho \right]^{2/3}\,,\label{vhe}
 \ee
the RS regime,
 \be
 m_\alpha^4 \gg
\rho\gg\sigma \equiv m_\sigma^4 ~ \Rightarrow ~ H^2\approx
{\kappa_4^2 \over 6\sigma}\, \rho^{2}\,,\label{he}
 \ee
the 4D regime,
 \be
\rho\ll\sigma~ \Rightarrow ~ H^2\approx {\kappa_4^2 \over 3}\,
\rho\,. \label{gr}
 \ee
The GB regime, when the GB term dominates gravity at the highest
energies, above the brane tension, can usefully be characterized
as
 \be\label{gbr}
H^2\gg \alpha^{-1} \gg \mu^2\,,~~ H^2 \propto \rho^{2/3}\,.
 \ee

The brane energy density should be limited by the quantum gravity
limit, $\rho<M_5^4$, in the high-energy regime. By
Eq.~(\ref{vhe}),
 \be
\rho<M_5^4 ~\Rightarrow~ H < \left({\pi M_5 \over 2\alpha}
\right)^{1/3}.
 \ee
In addition, since $\rho \gg m_\alpha^4$, we have~\cite{footnote}
 \be
\label{alphagreater} M_5 \gg m_\alpha ~\Rightarrow~ \alpha \gg {2
\over (8\pi M_5)^2}\,.
 \ee
Combining these two equations leads to
 \be\label{hlim}
m_\alpha^4\ll \rho < M_5^4 ~\Rightarrow~ H \ll 4\pi^{3/2}\, M_5\,.
 \ee
Comparing Eqs.~(\ref{alphagreater}) and (\ref{lim2}), we also find
that
 \be
\ell \gg {1\over 8\pi M_5}\,,
 \ee
which is equivalent to $M_4\gg M_5$, since, by Eq.~(\ref{k4k5}),
we have $\ell \approx (M_4/M_5)^2M_5^{-1}$.

\section{Bulk metric perturbations}

A de Sitter brane in an AdS$_5$ bulk is a solution to the GB field
and junction equations. The brane has a constant Hubble rate $H$,
and hence constant energy density $\rho>0$, which is added to the
brane tension $\sigma$, thus effectively breaking the RS
fine-tuning, as can be seen by comparing Eq.~(\ref{mf}) with
Eq.~(\ref{sig'}). Inflation in the extreme slow-roll regime may be
modelled by this solution.

The bulk metric satisfies Eq.~(\ref{bg}), and may be written in
the form
 \be
\label{bulksol} \vu d\bar{s}^2= A^2(z)\left[ \gamma_{\mu \nu}
dx^{\mu} dx^{\nu} + dz^2 \right]\,,
 \ee
where $\gamma_{\mu\nu}$ is the 4D de Sitter metric
($-dt^2+e^{2Ht}d\vec{x}^2$), and the conformal warp factor is
 \be \label{warp}
A(z)=\frac{H}{\mu\,\sinh Hz}\,,
 \ee
with $Z_2$ symmetry understood. The brane is at fixed position
$z=z_0>0$, which we can choose so that $A(z_0)=1$ (i.e., $\sinh
Hz_0=H/\mu$). The horizon is at $z=\infty$.

Consider now the 5D spin-2 metric perturbations, $\vu
\bar{g}_{ab}\to \,\vu \bar{g}_{ab}+\delta \,\vu {g}_{ab}$, where
$\delta \,\vu {g}_{ab}$ is 5D transverse traceless. For these
perturbations, Eq.~(\ref{5d}) shows that $\delta \H^a{}_b=0$, so
that the wave equation for the perturbations is
 \be
\delta \R^a{}_b=0\,,
 \ee
the same as in the RS case. This means that the bulk mode
solutions for metric perturbations will be the same as in the RS
case~\cite{lmw}, but the GB junction conditions will introduce
changes to the normalization and amplitudes of the modes, as
discussed below.

In the gauge $\delta \,\vu {g}_{az}=0$, we can write the perturbed
metric in the form
 \be
 \vu ds^2=\,\vu d\bar{s}^2+A^{1/2}h_{\mu\nu} dx^{\mu}
dx^{\nu}\,,
 \ee
where the factor $A^{1/2}$ is introduced for later convenience.
The perturbation may be decomposed into Kaluza-Klein (KK) modes,
 \[
h_{\mu\nu}(x,z) \to h_{\mu \nu}^\m(x) \phi_m(z)\,,
 \]
where integration (respectively sum) over the continuous
(respectively discrete) modes is understood, and
 \be
\gamma^{\mu \nu} h^\m_{\mu \nu}=0=\nabla^{\mu} h^\m_{\mu \nu} \,.
 \ee
Here and below, $\nabla_\mu$ denotes the covariant derivative of
the de Sitter metric $\gamma_{\mu\nu}$.

Because the brane is maximally symmetric, the wave equation
separates in brane-based coordinates (as in the case
$\alpha=0$~\cite{lmw}). The 4D part of the wave equation is
(compare~\cite{deser,fk})
 \be \label{4Dwe}
\Box \,h^\m_{\mu \nu} - 2 H^2 h^\m_{\mu \nu} = m^2 h^\m_{\mu
\nu}\,,
 \ee
which describes the propagation of 4D massive modes on a de Sitter
background. Here $\Box=\nabla^\mu\nabla_\mu$. The spin-2 quantity
$h^\m_{\mu\nu}$ encodes the 5 polarizations of the 5D graviton.
This corresponds, from the viewpoint of a 4D observer, to 2
polarizations in a 4D tensor mode, 2 polarizations in a 4D vector
mode (gravi-vector or gravi-photon), and 1 polarization in a 4D
scalar mode (gravi-scalar). Each of these will have in general a
zero-mode, i.e., a massless mode on the brane, and the massless
modes satisfy the same junction condition as in the RS case (see
below). However, for a single de Sitter brane, the zero mode
perturbation $h^{(0)}_{\mu\nu}$ has only 2 independent degrees of
freedom, corresponding to the usual 4D graviton. There are {\em
no} massless modes for the gravi-vector and
gravi-scalar~\cite{tanmon,gens}; these degrees of freedom can be
set to zero by the remaining gauge freedom on the brane~\cite{fk}.

The massive scalar and vector modes by contrast are not
degenerate. They have the same behaviour in the bulk as the
massive tensor modes. The massive modes of the 4D tensor
perturbations satisfy the same bulk wave equation as in the RS
case, but the junction condition at the brane is very different.
The 4D tensor part of Eq.~(\ref{gjc}) gives
\begin{equation}
\delta\! K^\mu{}_\nu \propto~\alpha \,\delta\! R^\mu{}_\nu .
\label{junc}
\end{equation}
In the RS case $\alpha=0$, the right-hand side of this equation is
zero, leading to the Neumann boundary condition,
$(A^{-3/2}\phi_m)'(z_0)=0$. On the brane, the perturbed Ricci
tensor is given by~\cite{lmw}
 \be
2\,\delta\! R_\mu{}^\nu=\ddot{h}_\mu{}^\nu + 3H
\dot{h}_\mu{}^\nu-e^{-2Ht}\partial_i\partial^i\, h_\mu{}^\nu \,.
 \ee
Separating variables, it follows that $\delta R_\mu{}^\nu\propto
m^2h_\mu{}^\nu$. Thus Eq.~(\ref{junc}) shows that in the GB case,
the boundary condition is of the form
 \be\label{junc2}
(A^{-3/2}\phi_m)'(z_0) \propto \alpha m^2 \phi_m(z_0)\,.
 \ee
The precise form of the boundary condition is given in
Eq.~(\ref{jc}) below.

\section{4D tensor perturbations}

The wave equation for the massive tensor modes can be written in
the form
 \be \label{zwe}
-\D_+ \left[ q(z) \, \D_- \, \phi_m(z)\right] = m^2 \, w(z) \,
\phi_m(z)\,,
 \ee
where we define the operators
 \be \label{D}
\D_{\pm}=\frac{d}{dz} \pm \frac{3}{2}\,\frac{A'}{A}\,,
 \ee
and the factors
 \bea  \label{q}
q&=&1-4\alpha\,A^{-4}\,\left(A'^2-A^2 H^2\right)\,,\\
w &=&1-4\alpha\,A^{-4}\,\left(A A''-A'^2\right)\,.
 \eea
This form of the wave equation explicitly incorporates the GB
junction condition. By Eq.~(\ref{warp}),
 \be\label{w}
w=1-4\alpha\mu^2-4\alpha A^{-3}[A']\delta(z-z_0)\,,
 \ee
where $[A']=2A'(z_0^+)=-2\sqrt{\mu^2+H^2}$ is the jump in $A'$
across the brane. Note from Eqs.~(\ref{warp}), (\ref{q}) and
(\ref{w}) that, for $z\neq z_0$, $q=w=1-4\alpha\mu^2$. Thus for
$z>z_0$, Eq.~(\ref{zwe}) reduces to the Schr\"odinger-type
equation,
 \bea
-\D_+\D_-\phi_m &=& -\phi''_m + \bigg[\frac{15}{4}\,
\frac{H^2}{\sinh^2 Hz}+\frac{9}{4}H^2\bigg]\phi_m \nonumber\\
{}&=&m^2\,\phi_m \,, \label{schrod}
 \eea
exactly as in the RS case $\alpha=0$~\cite{gs,lmw}.

The boundary condition for $\phi_m$ at $z=z_0$ is
 \be \label{jc}
\D_-\;\phi_m(z_0^+)= - \alpha m^2
\left[\frac{4\sqrt{\mu^2+H^2}}{1-4\alpha\mu^2}\right]
\phi_m(z_0)\,,
 \ee
and is of the form given in Eq.~(\ref{junc2}). This may be
obtained by matching the distributional parts of Eq.~(\ref{zwe}).
It is important to note that this boundary condition depends on
the mass of the modes, $m^2$, due to the $\alpha$-corrections (the
zero-mode, $m=0$, has the same boundary condition as in the RS
case). As a result, the scalar product of the eigenmodes
functional space has to include suitable boundary
terms~\cite{cdd}. It may be checked that the eigenmodes resulting
from Eqs.~(\ref{schrod}) and (\ref{jc}) are orthogonal with
respect to the following scalar product:
 \bea \label{orthonorm}
(\phi_m , \phi_n) = 2(1-4\alpha \mu^2) \int_{z_0}^{\infty}\! dz \,
\phi_m(z)\,\phi_n(z) \nonumber
\\ + 8\alpha\,\sqrt{\mu^2+H^2}\;\phi_m(z_0)\,\phi_n(z_0) =
\delta(m,n) \,.
 \eea
The normalization in the last equality denotes a Kroneker symbol
for the discrete modes and a Dirac distribution for the continuous
ones. Note that Eq.~(\ref{orthonorm}) reduces formally to
\[
(\phi_m , \phi_n) =\int_{\mathrm{bulk}} dz \,w \phi_m \phi_n\,,
\]
when $Z_2$-symmetry is imposed and the boundary term in
Eq.~(\ref{w}) is taken into account. When $\alpha\ge 0$, the norm
of the modes $|\!|\phi_m|\!|^2=(\phi_m,\phi_m)$ is always positive
for the branch of solutions chosen in Eq.~(\ref{ell}) and it
reduces to the usual norm for $\alpha=0$.

With the orthonormal conditions in Eq.~(\ref{orthonorm}), the
effective action for the metric perturbation, to second order, is:
 \bea
S &=& \frac{1}{2\kappa_5^2}\bigg\{\frac{1}{4}\int
d^4x\sqrt{-\gamma} \left[h^{\m \mu \nu}\, \Box h^\m_{\mu \nu}
\right.
\nonumber\\
&&\left.~{} -2H^2 h^{\m\mu \nu} h^\m_{\mu \nu} -m^2 h^{\m\mu \nu}
h^\m_{\mu \nu} \right] \bigg\}\,\, , \label{Seff}
 \eea
where the term in braces is the standard one for 4D (massive)
gravitons on a de Sitter background.

We now consider the spectrum of modes resulting from
Eqs.~(\ref{schrod}) and (\ref{jc}). There is a normalizable
bound-state zero-mode, as in the RS case:
 \be \label{zero}
\phi_0(z)=C\,A^{3/2}(z)\,,
 \ee
where the real constant $C$ will be determined in the following.
The asymptotic value of the Schrodinger potential in
Eq.~(\ref{schrod}), i.e., $\frac{9}{4}H^2$, gives the threshold
between the discrete and continuous spectra: $m^2>\frac{9}{4}H^2$,
as in the RS case~\cite{lmw,gs}. For the massive modes in the
continuous tower, the two linearly independent solutions of
Eq.~(\ref{schrod}) oscillate with constant amplitude for
$z\rightarrow\infty$. The boundary condition Eq.~(\ref{jc}) gives
$\phi_m(z)$ as a particular combination of these two solutions,
for every $m$. These modes are normalizable as plane waves and
form the continuous spectrum of Eqs.~(\ref{schrod}) and
(\ref{jc}).

For $m^2<\frac{9}{4}H^2$ on the other hand, Eq.~(\ref{schrod})
admits only one independent normalizable solution for each $m$.
The corresponding mode behaves as a decreasing exponential for
$z\rightarrow\infty$. For $\alpha=0$, the only such mode which
satisfies the junction condition is the massless mode,
Eq.~(\ref{zero}). In GB gravity however, this issue is more subtle
because of the explicit dependence of the boundary condition
Eq.~(\ref{jc}) on $m^2$.

In order to see whether the junction conditions allow for discrete
states other than the zero mode, it is convenient to introduce the
new modes:
 \be \label{Fi}
\Phi_m(z)=\D_-\;\phi_m(z)\,,
 \ee
which are the partners of the modes $\phi_m$ in super-symmetric
quantum mechanics~\cite{cks}. They have the same spectrum except
for the zero-mode: $\Phi_0$ vanishes identically, by
Eqs.~(\ref{D}) and (\ref{zero}). The wave equation for $\Phi_m$ is
found by applying $\D_-$ to Eq.~(\ref{schrod}):
 \bea  -\D_-\D_+\Phi_m &=& -\Phi_m'' +
\bigg[\frac{3}{4}\,\frac{H^2}{\sinh^2 Hz}+\frac{9}{4}H^2\bigg]
\Phi_m \nonumber\\ &=& m^2\,\Phi_m \,. \label{schrod2}
 \eea
The boundary condition follows from Eqs.~(\ref{schrod}) and
(\ref{jc}),
 \be \label{jc2}
\Phi_m'(z_0^+)=\bigg[\frac{1-4\alpha\mu^2}
{4\alpha\sqrt{\mu^2+H^2}} + \frac{3}{2}\sqrt{\mu^2+H^2}\bigg]
\Phi_m(z_0)\,,
 \ee
for $\alpha\neq 0$, while $\Phi_m(z_0)=0$ for $\alpha=0$.

In particular, this boundary condition no longer involves the mass
of the modes (and reduces to Dirichlet-type for $\alpha=0$). This
is the essential property we need. Multiplying Eq.~(\ref{schrod2})
by $\Phi_m$ and integrating by parts, we find that
 \bea
&& \left(m^2-\frac{9}{4}H^2\right)\int_{z_0}^{\infty}\! dz\,
\Phi_m^2 = \frac{3}{4}H^2\int_{z_0}^{\infty}\! dz\,
\frac{\Phi_m^2} {\sinh^2Hz} \nonumber\\ &&~~~{}
 +  \int_{z_0}^{\infty}\! dz\,\Phi_m'^2
+ \bigg[-\Phi_m\Phi_m'\bigg]_{z_0}^{\infty}. \label{proof}
 \eea
Consider now a would-be normalizable mode $\phi_m$ with
$m<{3\over2}H$. Its partner $\Phi_m$ must decrease exponentially
when $z\rightarrow\infty$, as does $\phi_m$. The corresponding
upper boundary term at infinity in Eq.~(\ref{proof}) therefore
vanishes. The lower boundary term on the brane, by
Eq.~(\ref{jc2}), is positive for the minus branch of solutions,
defined in Eq.~(\ref{ell}), and for $\alpha\geq 0$. It vanishes
for $\alpha=0$. Thus in this case, the right-hand side of
Eq.~(\ref{proof}) is positive, while the left-hand side is
negative. This can be satisfied only for $\Phi_m=0$, i.e., for
$m=0$. We therefore conclude that the spectrum of KK modes
consists only of \begin{itemize}\item  the massless bound-state
zero-mode, \item a continuum of states with $m>{3\over2}H$,
\end{itemize} as in the RS case $\alpha=0$~\cite{gs,lmw}.

This feature is crucial for discussing stability issues as well as
the gravitational waves produced along the brane. In particular,
the spectrum rules out the existence of 4D massive gravitons with
$m^2<2H^2$ in Eq.~(\ref{4Dwe}), which would have signalled a
classical instability of the model~\cite{deser} (see also
Ref.~\cite{cho} for a recent discussion in the braneworld
context). It has been shown that a mass gap for de Sitter branes
is quite generic in Einstein gravity~\cite{fk}. In particular it
still holds if a second $Z_2$-symmetric brane is introduced in the
background, Eq.~(\ref{bulksol}), say at $z=z_2>z_0$. We just note
here however that we can {\em not} reach the same conclusion in
Gauss-Bonnet gravity. In particular, the argument above would fail
in this case, because the new boundary term at $z=z_2<\infty$ in
Eq.~(\ref{proof}) would then be negative (while it still vanishes
for $\alpha=0$). In fact, if we solve Eq.~(\ref{schrod})
explicitly and impose Eq.~(\ref{jc}), we can show that tachyonic
modes with $m^2<0$ ($<2H^2$) {\em may exist} for the 2-brane
system with $\alpha>0$ (as well as for the 1-brane case with
$\alpha<0$). This system may therefore suffer from the same spin-2
tachyonic instability present for Minkowski branes~\cite{cd}.
(Note that the tachyonic instability in the case of two de Sitter
branes with Einstein gravity~\cite{gens,fk2} is a spin-0 radion
mode.)

\section{Amplitude of the zero-mode}

We can now estimate the spectrum of graviton fluctuations
generated in de Sitter inflation on the brane, by treating each
mode as a quantum field in four dimensions, as in the RS
case~\cite{lmw,fk} (see Refs.~\cite{grs,kkt} for a
five-dimensional approach).

For $m^2>{9\over4}H^2$, the massive modes are strongly suppressed
on large scales and remain in their vacuum state~\cite{lmw,fk}.
These modes can therefore be neglected in the following. However,
the zero-mode is over-damped and acquires a spectrum of classical
perturbations on super-horizon scales. For $m^2=0$, the effective
action Eq.~(\ref{Seff}) has the standard form of 4D general
relativity, except for the overall factor $\kappa_5^2$ instead of
$\kappa_4^2$, which rescales the amplitude of quantum fluctuations
in $h^{(0)}_{\mu \nu}$ accordingly~\cite{lmw}. Thus the amplitude
of gravitational waves produced on super-horizon scales on the
brane is given by
 \be \label{AT}
A_T^2= \kappa_5^2 \;\phi_0^2(z_0)\bigg(\frac{H}{2\pi}\bigg)^2.
 \ee
The normalization of the discrete zero-mode, $\phi_0(z_0)=C$,
introduces further rescaling relative to the 4D result. By
Eqs.~(\ref{orthonorm}) and (\ref{zero}), the condition
$(\phi_0,\phi_0)=1$ gives:
 \be \label{psi0}
C^{-2}=\frac{(1+4\alpha\mu^2)}{\mu}\;F_{\alpha}^{-2}(H/\mu)\,,
 \ee
where we used Eq.~(\ref{k4k5}), and where
 \be
F_{\alpha}^{-2}(x)  =\sqrt{1+x^2} -\left(\!\frac{1-4\alpha\mu^2}
{1+4\alpha\mu^2}\!\right) x^2\sinh^{-1}{1\over x}\,. \label{F}
 \ee
This generalizes the function $F_0(x)$ found for the RS
case~\cite{lmw}. When $x \equiv H/ \mu\to 0$, we have
$F_{\alpha}\to 1$; the amplitude of the normalized zero-mode on a
Minkowski brane measures the ratio between the effective 4D Newton
constant $\kappa^2_4$, and the 5D constant $\kappa^2_5$.

The modified tensor amplitude is therefore
 \be \label{tensoramplitude}
A_T^2= \kappa_4^2 \bigg(\frac{H}{2\pi}\bigg)^2
F_{\alpha}^2(H/\mu)\,,
 \ee
and the correction to standard 4D general relativity lies in the
last factor:
 \be
F_\alpha^2={A_T^2 \over [A_T^2]_{4\rm D}}\,.
 \ee
This correction depends on the GB coupling $\alpha$ and on the
energy scale at which inflation occurs, relative to the 5D
curvature scale $\mu$, and it reduces to the result of
Ref.~\cite{lmw} for the RS case $\alpha=0$. (The correction to the
4D result may also be expressed via an effective Planck mass
during inflation, following Ref.~\cite{fk}.)

The GB term introduces significant corrections to the RS case. In
the GB regime, as characterized by Eq.~(\ref{gbr}), we have
 \be
\label{Falpha} F_{\alpha}^2(H/\mu) \approx
\frac{(1+4\alpha\mu^2)}{8\alpha\mu^2} \left({H\over \mu}
\right)^{-1}\,,
 \ee
while the RS case $\alpha=0$ yields
 \be
\label{Fzero} F_{\alpha=0}^2(H/\mu) \approx
{3\over2}\,\frac{H}{\mu}\,.
 \ee
Thus the GB term {\em suppresses} tensor perturbations relative to
the 4D result, at high energies, unlike the RS case where the
tensor amplitude is strongly enhanced. If we consider the GB term
as a perturbative correction to the Einstein-Hilbert 5D action,
then $\beta\equiv 4\alpha\mu^2 \ll 1$, and there is an RS regime
as the energy drops (but remains above the brane tension). Thus we
expect that the tensor amplitude is enhanced at lower energies (RS
regime) and suppressed at higher energies (GB regime), so that
there is a maximum at intermediate energies. This qualitative
behaviour is confirmed in Figs.~1 and 2.


\begin{figure}\label{F1}
\begin{center}
\includegraphics[height=3.5in,width=3.5in,angle=-90]{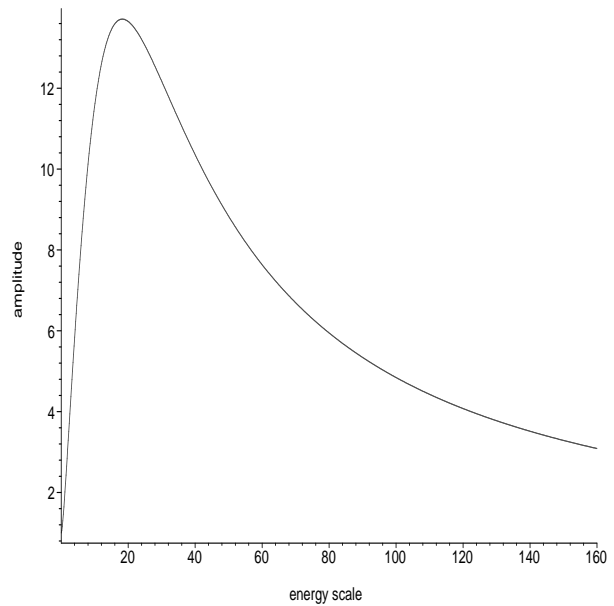}
\caption{The dimensionless amplitude $F_\alpha^2$ of the tensor
zero-mode relative to the 4D general relativity result, plotted
against the dimensionless energy scale of inflation, $H/\mu$. (The
Gauss-Bonnet coupling is given by $\beta\equiv 4 \alpha\mu^2
=10^{-3}$.)}
\end{center}
\end{figure}


\begin{figure}\label{F2}
\begin{center}
\includegraphics[height=3.5in,width=3.5in,angle=-90]{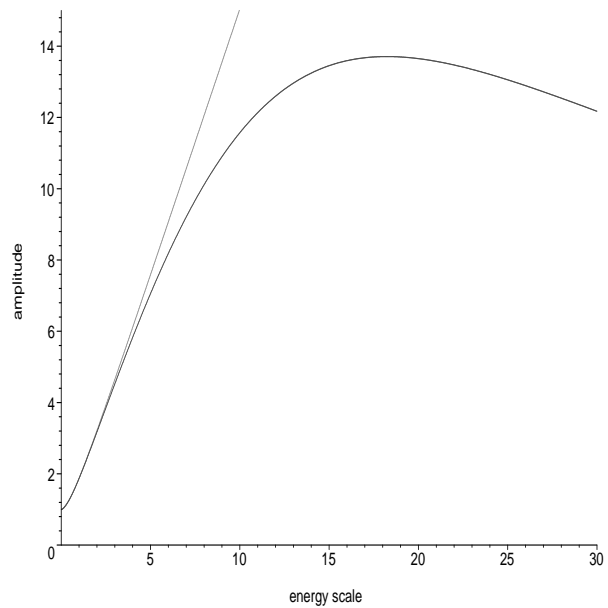}
\caption{As in Fig.~1, but with the upper curve showing the
Randall-Sundrum case $\alpha=0$.}
\end{center}
\end{figure}

The quantum gravity limit $\rho<M_5^4$ imposes an upper limit on
the energy scale $x$; by Eqs.~(\ref{k4k5}) and (\ref{hlim}),
 \be\label{xlim}
x\equiv {H\over \mu} \ll \left({M_4 \over M_5} \right)^2.
 \ee
There is a range of energies, $0<H<H_0$, where $H_0$ is the
solution to $F_\alpha^2(H_0)=1$, such that $F_\alpha^2(H)>1$,
i.e., such that the amplitude of gravitational waves from GB
inflation is greater than the standard 4D amplitude. By
Eqs.~(\ref{al}) and (\ref{Falpha}),
 \be
0< H < H_0 \approx {1\over 8\alpha\mu}~ \Rightarrow~
F_\alpha^2>1\,.
 \ee
Changing the value of $\alpha$ changes the location and height of
the maximum, and the value of $H_0$, but the maximum always has
$F_\alpha^2>1$. In all cases, $F_\alpha(0)=1$, and the asymptotic
behaviour as $x\to\infty$ is $F_\alpha^2 \sim x^{-1}$, as given by
Eq.~(\ref{Falpha}). The maximum of $F^2_{\alpha}$ increases as
$\alpha$ decreases, and so does the range $H_0$ of energies for
which the amplitude exceeds the 4D result. In the RS limit
$\alpha\to 0$, there is no maximum: $F_0^2$ is monotonically
increasing without bound for $x\to\infty$, as illustrated in
Fig.~2.

The maximum gravitational-wave amplitude relative to the standard
4D result for $\alpha>0$ is given by
 \be
(F_\alpha^2)_{\rm max}={A^2_{T,{\rm max}} \over [A_T^2]_{4\rm D}}
= {(1+4\alpha\mu^2)\sqrt{1+H_{\rm m}^2/\mu^2} \over 1+4\alpha\mu^2
+4\alpha H_{\rm m}^2}\,,
 \ee
where the critical inflation energy scale $H_{\rm m}$ is
determined by the root of the equation
 \be
\sqrt{1+x_{\rm m}^2}\,\sinh^{-1}{1\over x_{\rm m}}={1\over
1-4\alpha\mu^2}\,.
 \ee

\section{The tensor to scalar ratio}

It is well known that in standard, slow-roll inflation driven by a
single inflaton field, the scalar and tensor perturbations are not
independent, but are instead related by a consistency relation.
(For a review, see, e.g., Ref.~\cite{llkcba}). To lowest order in
the slow-roll approximation, the ratio of the tensor to scalar
perturbations is given by
\begin{equation}
\label{standardconsistent} \frac{A_T^2}{A_S^2} = -
\frac{1}{2}n_T\,,
\end{equation}
where $n_T \equiv d \ln A_T^2 /d \ln k$ represents the tilt of the
tensorial spectrum and $k$ is comoving wavenumber. An identical
relation holds in 4D scalar-tensor and other generalized Einstein
theories~\cite{tg}, and also in the RS scenario~\cite{hl1,hl2},
and in a 5D braneworld model where the radion field is
stabilized~\cite{gklr}. Formally, the degeneracy in the braneworld
models arises because the function that parametrizes the
corrections to the gravitational wave amplitude satisfies a
particular first-order differential equation~\cite{hl1,st}.

Given the potential importance of the consistency relation as a
way of reducing the number of independent inflationary parameters,
and of testing the inflationary scenario, it is important to
investigate whether the above degeneracy is lifted when GB effects
are included in the bulk action as a correction to the RS model.
Furthermore, the relative contribution of tensor perturbations to
CMB anisotropies is also an important quantity for constraining
inflationary models~\cite{liddle}, and we will consider how the GB
term affects this. In the RS case, although both tensor and scalar
perturbations are enhanced, the tensors are enhanced less and thus
the relative tensor contribution is suppressed in comparison with
the standard case. First we need to compute the scalar
perturbation amplitude $A_S$.

\subsection{Scalar perturbations from GB brane inflation}

We assume that there is no scalar zero-mode contribution from bulk
metric perturbations (5D gravitons) during inflation, and that the
massive scalar KK modes may be neglected in inflation. The latter 
is true in the exact de Sitter inflation case, as discussed above.
The scalar massive modes may be ignored, since they
are heavy ($m>{3\over2}H$) and stay in their vacuum state during
inflation, both for the RS case and the GB generalization. For
more general inflationary expansion, it may not be realistic to
ignore the massive modes, but in the extreme slow-roll limit, it
may be a reasonable approximation to neglect the bulk metric
perturbations. In this approximation we can take over the standard
4D results that do not depend on the standard Friedman equation,
as in the RS case~\cite{mwbh}.

Conservation of energy-momentum on the brane, Eq.~(\ref{tmunu}),
implies that the adiabatic matter curvature perturbation $\zeta$ on a uniform
density hypersurface is conserved on large scales, independently
of the gravitational physics~\cite{wmll}. Consequently, the
amplitude of a given mode that re-enters the Hubble radius after
inflation is given by $A_S^2 = H^4/(2\pi^2 \dot{\varphi}^2 )$.
Here and in similar expressions in this Section, equality is to be
understood as equality at the lowest order in the slow-roll
approximation. (The normalization is chosen such that $A_S^2 = 2
\langle \zeta^2 \rangle$.) In this limit, the scalar field
equation, $\dot{\varphi} = -V'(\varphi)/3H$, implies that the
amplitude of scalar (density) perturbations is given by
\begin{equation}
\label{scalaramplitude} A_S^2 = \frac{9}{2 \pi^2}
\frac{H^6}{V'^2}\,.
\end{equation}
Using Eqs.~(\ref{mfe}), (\ref{chi}), (\ref{sig'}) and
(\ref{k4k5}), we can write this in terms of the standard result,
$[A_S^2]_{4\rm D}=\kappa_4^6 V^3/ 6\pi^2 V'^2$, as follows:
 \be
A_S^2 = [A_S^2]_{4\rm D}\, G^2_\alpha(H/\mu)\,,
 \ee
where
 \be \label{gg}
G^2_\alpha(x)=\left[{3(1+\beta) x^2 \over 2\sqrt{1+x^2}(3-\beta
+2\beta x^2)+2(\beta-3)}\right]^{3},
 \ee
with $\beta \equiv 4\alpha\mu^2$.

The scalar spectral index, $n_S-1 \equiv d \ln A_S^2/d\ln k |_{k=aH}$
can be expressed in terms of the slow--roll parameters, 
$\epsilon \equiv -\dot{H}/H^2$ and $\eta \equiv V''/3H^2$, 
such that \cite{tsm}
\begin{equation}
n_S-1 = -6\epsilon +2 \eta
\end{equation}
where 
\begin{eqnarray}
\frac{\epsilon}{\epsilon_{\rm RS}} = \frac{2(1-\beta 
)^4 \sinh \frac{2}{3}
\chi {\rm tanh} \, \chi \sinh^2 \chi}{9(1+\beta ) (3 -\beta )
[(1-\beta ) \cosh \frac{2}{3} \chi -1]^2} \, , 
\\
\frac{\eta}{\eta_{\rm RS}} = \frac{2 (1-\beta )^3 \sinh^2 \chi}{3(1+\beta 
)(3-\beta ) [ (1-\beta ) \cosh \frac{2}{3} \chi -1]} \, , 
\end{eqnarray}
and $\epsilon_{\rm RS} \equiv 2\sigma V'^2/(\kappa_4^2 V^3)$ and 
$\eta_{\rm RS} \equiv 2\sigma V''/(\kappa^2_4V^2)$ 
are the corresponding RS slow--roll parameters \cite{mwbh}.

As in the case of tensor perturbations, the scalar perturbations
with GB corrections behave very differently compared to the RS
case. At high energies, the GB term again leads to a suppression
of scalar perturbations relative to the standard result.  In the
GB regime, as characterized by Eq.~(\ref{gbr}), we have
 \be \label{ggh}
G^2_\alpha \approx {27\over64}\left({1+\beta \over \beta}
\right)^3 {1\over x^3}\,.
 \ee
By contrast, in the RS case, scalar perturbations are strongly
enhanced at high energies. Thus we have a similar qualitative
behaviour to the tensor case: there is an RS regime of
amplification at lower energies, and a GB regime of suppression,
with a maximum at intermediate energies. The qualitative behaviour
of the dimensionless amplitude of scalar perturbations is shown in
Fig.~3.


\begin{figure}\label{F3}
\begin{center}
\includegraphics[height=3.5in,width=3.5in,angle=-90]{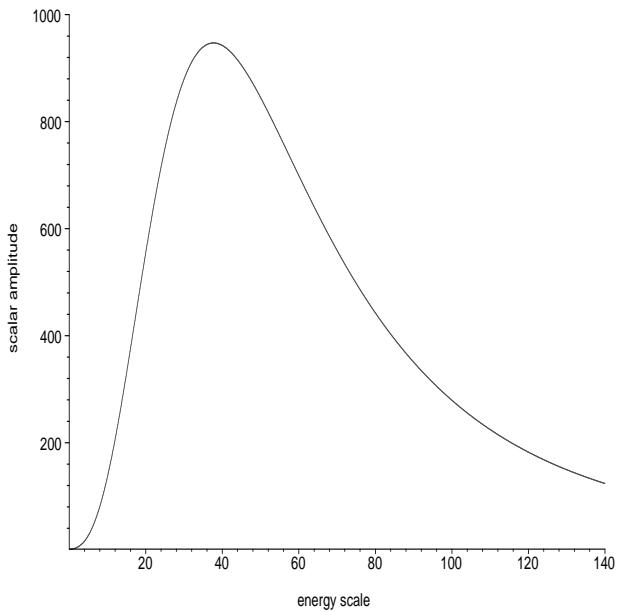}
\caption{The dimensionless amplitude $G_\alpha^2$ of density
perturbations relative to the 4D general relativity result,
plotted against the dimensionless energy scale of inflation,
$H/\mu$. (The Gauss-Bonnet coupling is given by $\beta\equiv 4
\alpha\mu^2=10^{-3}$.)}
\end{center}
\end{figure}


\begin{figure}\label{F4}
\begin{center}
\includegraphics[height=3.5in,width=3.5in,angle=-90]{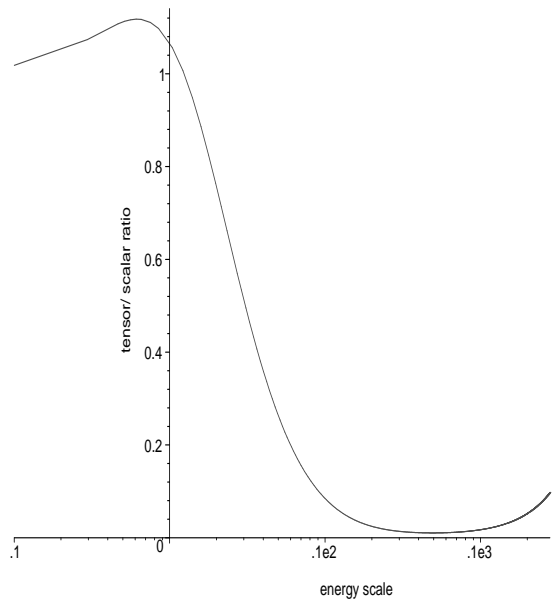}
\caption{The ratio of tensor to scalar perturbations,
$R=A_T^2/A_S^2$, relative to the 4D general relativity ratio,
plotted against the dimensionless energy scale of inflation, $\log
(H/\mu)$. (The Gauss-Bonnet coupling is given by $\beta\equiv 4
\alpha\mu^2=10^{-3}$.)}
\end{center}
\end{figure}

Furthermore, the different scaling of the scalar and tensor
amplitudes at high energies leads to another intriguing difference
from the RS case. In the RS case, tensors are enhanced less
strongly than scalars, so that the tensor/ scalar ratio
$R=A_T^2/A_S^2$ is suppressed in comparison with the standard
case. When there is a GB term, Eqs.~(\ref{Falpha}) and (\ref{ggh})
show that the scalars are more strongly suppressed at high
energies than the tensors, so that the tensor/ scalar ratio is
{\em enhanced} at high energies. At lower and intermediate
energies the ratio is a complicated function of $x$, given the
interplay between RS and GB effects. It follows from
Eqs.~(\ref{F}) and (\ref{gg}) that
 \bea \label{ratio}
&&{R \over R_{4\rm D}}= \\&&{}{[2\sqrt{1+x^2}(3-\beta +2\beta
x^2)+2(\beta-3)]^3\over 27(1+\beta)^2x^6[(1+\beta)\!
\sqrt{1+x^2}-(1-\beta) x^2\sinh^{-\!1}x^{-\!1}]}\!,\nonumber
 \eea
where $R_{4\rm D}=[A_T^2/A_S^2]_{4\rm D}$. The ratio of tensor to
scalar amplitudes, relative to the standard ratio, has a maximum
at low energies, a minimum at high energies, and grows like $x^2$
at very high energies. This is illustrated in Fig.~4.

\subsection{Consistency relation}

The consistency relation for the GB braneworld is derived by
differentiating the gravitational wave amplitude,
Eq.~(\ref{tensoramplitude}), with respect to comoving wavenumber
$k (\varphi)  =a (\varphi) H (\varphi)$. In the extreme slow-roll
limit, variations in the Hubble parameter are negligible relative
to changes in the scale factor. This implies that the tensor
spectral index can be expressed as
\begin{equation}
\label{tensortilt} n_T =- \frac{d \ln  \left( xF_{\alpha}
\right)^{-2}}{d \ln x}\,{a\over H}\, \frac{d H}{d a}  .
\end{equation}

The GB braneworld correction to the gravitational amplitude,
Eq.~(\ref{F}), satisfies an important first-order differential
equation:
\begin{equation}
\label{importantode} \frac{d}{d \ln x} \left[ \ln
(xF_{\alpha})^{-2} \right] = -\frac{2F^2_{\alpha}[1+\beta (1+x^2)]
}{(1+\beta)\sqrt{1+x^2}} \,.
\end{equation}
Furthermore, the scalar field equation can be expressed in the
form
\begin{equation}
\label{scalareom} a \frac{dH}{d a} = - \frac{dH}{dV}
\frac{V'^2}{3H^2}\,.
\end{equation}
Hence, substituting Eqs.~(\ref{tensoramplitude}),
(\ref{scalaramplitude}), (\ref{importantode}) and
(\ref{scalareom}) into Eq. (\ref{tensortilt}) implies that the
tensor to scalar ratio is given by
\begin{equation}
\frac{A_T^2}{A_S^2}=-{Q \over 2}n_T \,,
\end{equation}
where
\begin{equation}
\label{defQ} Q^{-1} = \frac{6}{\kappa_4^2} \frac{[1+\beta
(1+x^2)]}{(1+\beta)\sqrt{1+x^2}} \, H \frac{dH}{dV}  \,.
\end{equation}
The function $Q(H)$ determines to what extent the degeneracy of
the consistency equation is lifted in GB braneworld inflation and
we therefore refer to it as the ``degeneracy factor". For our
normalization conventions, it takes the value $Q =1$ in the
standard and also the RS inflationary scenarios.

Differentiating Eqs.~(\ref{mfe}) and (\ref{chi}) with respect to
$\chi$, Eq. (\ref{defQ}) then implies that
\begin{eqnarray}
\label{Qresult} Q= {(1-\beta)\cosh \chi \over \left[1  + 2
(1-\beta) \sinh^2 ( {\chi}/{3})  \right] \cosh (\chi /3)} ,
\end{eqnarray}
where the result has been simplified by employing
Eq.~(\ref{k4k5}). Using the identity $\cosh \chi + \cosh (\chi /3)
= 2  \cosh (2 \chi /3) \cosh (\chi /3)$, we find that the
degeneracy factor takes the simple form:
\begin{equation}
\label{Qx} Q ={ 1+\beta + 2\beta x^2 \over 1+\beta + \beta x^2}
\,.
\end{equation}


\begin{figure}\label{F5}
\begin{center}
\includegraphics[height=3.5in,width=3.5in,angle=-90]{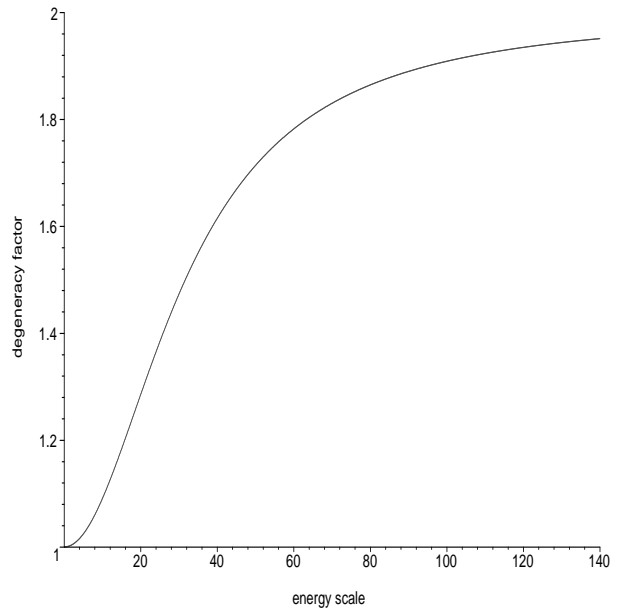}
\caption{The tensor/ scalar degeneracy factor $Q$, plotted against
the dimensionless energy scale of inflation, $H/\mu$. (The
Gauss-Bonnet coupling is given by $\beta\equiv 4
\alpha\mu^2=10^{-3}$.)}
\end{center}
\end{figure}

We may conclude immediately from Eq.~(\ref{Qx}) that GB effects
{\em lift} the degeneracy of the consistency equation, since the
factor $Q$ depends directly on the energy scale, $x$,
corresponding to the time when the observable modes went beyond
the Hubble radius during inflation. The standard form of the
consistency equation ($Q=1$) is recovered for $\alpha=0$, and also
in the limit $\beta x^2\equiv 4\alpha\mu^2x^2 \ll 1$,
corresponding to the regimes of Eqs.~(\ref{he}) and (\ref{gr}) in
the history of the GB braneworld. However, in the GB regime of
Eq.~(\ref{gbr}), the asymptotic form of the degeneracy factor is
$Q \rightarrow 2$. In this limit the consistency equation is given
by
\begin{equation}
\label{GBconsistent} \frac{A_T^2}{A_S^2} = -n_T  .
\end{equation}
The behaviour of $Q$ as a function of energy scale is shown in
Fig.~5.

It is interesting that Eq.~(\ref{GBconsistent}) is independent not
only of the specific form of the inflaton potential, but also of
the parameters of the model, specifically the brane tension,
$\sigma$, the bulk cosmological constant, $\Lambda_5$, and the GB
parameter, $\alpha$. As in the standard 4D and the 5D RS
scenarios, the consistency relation can be expressed entirely in
terms of observable parameters and, in this sense,
Eq.~(\ref{GBconsistent}) may be viewed as a model-independent
observable signature of GB braneworld inflation in the high energy
limit, Eq.~(\ref{vhe}).

\section{Conclusions}

Brane inflation offers a phenomenology that allows us to explore
some of the cosmological implications of ideas arising from string
and M~theory. The effects on inflationary perturbations from the
extra dimensional nature of gravity introduce new features that
need to be computed and then subjected to the constraints from
high-precision cosmological data. Here we have concentrated on
computing the corrections to the standard results for tensor and
scalar perturbations that are generated during slow-roll inflation
at energies where brane effects become dominant. This has
previously been done for the Randall-Sundrum braneworld, based on
5-dimensional Einstein gravity. We have introduced a Gauss-Bonnet
term, since string theory arguments indicate that this term is a
high-energy perturbative correction to the gravitational action.
This correction leads to significant qualitative changes, even in
the perturbative regime $\beta\equiv 4\alpha\mu^2 \ll 1$.

For the tensor perturbations, we have given an exact analysis,
including the 5D modes. The wave equation and its fundamental
solutions are not changed by the GB term. The spectrum contains a
normalizable zero-mode and a continuous tower of massive modes
after a mass gap, $m>{3\over 2}H$, as in the RS case. The massive
modes are not excited during inflation, as in the RS case.
However, the GB term changes the boundary conditions at the brane,
and therefore changes the normalization of the zero-mode, as shown
by Eq.~(\ref{F}):
 \[
{A_T^2 \over [A_T^2]_{4\rm D}}= \left[\sqrt{1+x^2}
-\left(\!\frac{1-\beta} {1+\beta}\!\right) x^2\sinh^{-1}{1\over
x}\right]^{-2}\!.
 \]
This leads in the GB regime to a suppression of tensor
perturbations relative to the standard result, unlike the
enhancement that arises in the RS case $\beta=0$.

For the scalar perturbations, we used an approximation where bulk
perturbations decouple from the density perturbations. We showed
that the GB modifications to the Friedman equation lead to a
significant change from the RS case, as given by Eq.~(\ref{gg}):
 \begin{eqnarray*}
{A_S^2 \over [A_S^2]_{4\rm D}}= \left[{3(1+\beta) x^2 \over
2\sqrt{1+x^2}(3-\beta +2\beta x^2)+2(\beta-3)}\right]^{3}\!.
 \end{eqnarray*}
These perturbations are again suppressed in the GB regime relative
to the standard result, unlike the RS enhancement.

Because the scalar suppression is stronger than the tensor
suppression in the GB regime, the relative tensor contribution, as
a fraction of the scalar amplitude, $R=A_T^2/A_S^2$, is enhanced
in the GB regime, in comparison with the standard result. This is
shown by Eq.~(\ref{ratio}):
 \begin{eqnarray*}
&&{R \over R_{4\rm D}}= \\&&{}{[2\sqrt{1+x^2}(3-\beta +2\beta
x^2)+2(\beta-3)]^3\over 27(1+\beta)^2x^6[(1+\beta)\!
\sqrt{1+x^2}-(1-\beta) x^2\sinh^{-\!1}x^{-\!1}]}\!.
 \end{eqnarray*}
By contrast, in the RS case the relative tensor contribution is
suppressed.

Furthermore, the consistency relation between the tensor/ scalar
ratio and the tensor spectral tilt is different in the GB case,
i.e.,
 \[
R=-\left({ 1+\beta + 2\beta x^2 \over 1+\beta + \beta x^2}
\right){n_T \over 2}\,,
 \]
by Eq.~(\ref{Qx}). The RS model by contrast has the same
consistency relation as the standard case, $R=-{1\over2}n_T$.

Our results provide a basis on which to confront the GB braneworld
with observational constraints, and this is under investigation.

\[ \]
{\bf Acknowledgements:}

We thank Pierre Binetruy, Christos Charmousis, Naresh Dadhich,
Stephen Davis, Lev Kofman, David Langlois, Jihad Mourad, Sergei
Odintsov, Renaud Parentani, Danielle Steer and David Wands for
helpful discussions. The work of RM is supported by PPARC.


\begin{thebibliography}{99}


\bibitem{RSII}
L. Randall and R. Sundrum, Phys. Rev. Lett. {\bf 83}, 4690 (1999)
[hep-th/9906064].

\bibitem{royreview}
P. Brax and C. van de Bruck, Class. Quantum Grav. {\bf 20}, R201
(2003) [hep-th/0303095]; R. Maartens, gr-qc/0312059; P. Brax, C.
van de Bruck, and A-C. Davis, hep-th/0404011.

\bibitem{adscft}
J. M. Maldacena, Adv. Theor. Math. Phys. {\bf 2}, 231 (1998)
[hep-th/9711200]; E. Witten, Adv. Theor. Math. Phys. {\bf 2}, 505
(1998) [hep-th/9803131]; S. Gubser, I. Klebanov, and A. Polyakov,
Phys. Lett. B{\bf 428}, 105 (1998) [hep-th/9802109]; O. Aharony,
S. Gubser, J. Maldacena, H. Ooguri, and Y. Oz, Phys. Rep. {\bf
323}, 183 (2000) [hep-th/9905111] .

\bibitem{hrr}
S. W. Hawking, T. Hertog, and H. Reall, Phys. Rev. D{\bf 62}
043501 (2000) [hep-th/0003052]; M. J. Duff and J. T. Liu, Phys.
Rev. Lett. {\bf 85}, 2052 (2000) [hep-th/0003237]; S. Nojiri, S.
D. Odintsov, and S. Zerbini, Phys. Rev. D{\bf 62}, 064006 (2000)
[hep-th/0001192]; S. Nojiri and S. Odintsov, Phys. Lett. B{\bf
484}, 119 (2000) [hep-th/0004097];  
S. Nojiri and S. D. Odintsov, J. High Energy Phys. 
{\bf 07}, 049 (2000) [hep-th/0006232]; 
L. Anchordoqui, C. Nunez, and
K. Olsen, J. High Energy Phys. {\bf 10}, 050 (2000)
[hep-th/0007064]; S. Nojiri and S. Odintsov, Phys. Lett. B{\bf
494}, 135 (2000) [hep-th/0008160]; S. Gubser, Phys. Rev. D{\bf
63}, 084017 (2001) [hep-th/9912001]; T. Shiromizu and D. Ida,
Phys. Rev. D{\bf 64}, 044015 (2001) [hep-th/0102035]; 
M. Cvetic, S. Nojiri and S. D. Odintsov,  
Nucl. Phys. {\bf B628}, 295 (2002) [hep-th/0112045]; 
J. E. Lidsey, S. Nojiri and S. D. Odintsov, 
J. High Energy Phys. {\bf 06}, 026 (2002) [hep-th/0202198].

\bibitem{largeN}
A. Fayyazuddin and M. Spalinski, Nucl. Phys. B{\bf 535}, 219
(1998) [hep-th/9805096]; O. Aharony, A. Fayyazuddin, and J.
Maldacena, J. High Energy Phys. {\bf 07}, 013 (1998)
[hep-th/9806159].

\bibitem{d86}
D. Lovelock, J. Math. Phys. {\bf 12}, 498 (1971); N. Deruelle and
J. Madore, Mod. Phys. Lett. A{\bf 1}, 237 (1986); N. Deruelle and
L. Farina-Busto, Phys. Rev. D{\bf 41}, 3696 (1990).

\bibitem{stringGB}
D. G. Boulware and S. Deser, Phys. Rev. Lett. {\bf 55}, 2656
(1985); B. Zwiebach, Phys. Lett. B{\bf 156}, 315 (1985); A. Sen,
Phys. Rev. Lett. {\bf 55}, 1846 (1985); R. R. Metsaev and A. A.
Tseytlin, Nucl. Phys. B{\bf 293}, 385 (1987).

\bibitem{onlylocal}
N. E. Mavromatos and J. Rizos, Phys. Rev. D{\bf 62}, 124004 (2000)
[hep-th/0008074]; I. P. Neupane, J. High Energy Phys. {\bf 09},
040 (2000) [hep-th/0008190]; I. P. Neupane, Phys. Lett. B{\bf
512}, 137 (2001) [hep-th/0104226]; K. A. Meissner and M.
Olechowski, Phys. Rev. Lett. {\bf 86}, 3708 (2001)
[hep-th/0009122]; Y. M. Cho, I. Neupane, and P. S. Wesson, Nucl.
Phys. B{\bf 621}, 388 (2002) [hep-th/0104227].

\bibitem{milder}
N. Deruelle and M. Sasaki, Prog. Theor. Phys. {\bf 110}, 441
(2003) [gr-qc/0306032].

\bibitem{wmap}
D. N. Spergel {\em et al.}, Astrophys. J. Suppl. {\bf 148}, 175
(2003) [astro-ph/0302209]; H. V. Peiris {\em et al.}, Astrophys.
J. Suppl. {\bf 148}, 213 (2003) [astro-ph/0302225].

\bibitem{gw}
A. A. Starobinsky, Pis'ma Zh. Eksp. Teor. Fiz. {\bf 30}, 719
(1979) (JETP Letters {\bf 30}, 682); L. F. Abbott and M. B. Wise,
Nucl. Phys. B{\bf 244}, 541 (1984).

\bibitem{Bmode}
R. Crittenden, J. R. Bond, R. L. Davis, G. Efstathiou, and P. J.
Steinhardt, Phys. Rev. Lett. {\bf 71}, 324 (1993)
[astro-ph/9303014]; M. Kamionkowski, A. Kosowsky, and A. Stebbins,
Phys. Rev. Lett. {\bf 78}, 2058 (1997) [astro-ph/9611125]; U.
Seljak and M. Zaldarriaga, Phys. Rev. Lett. {\bf 78}, 2054 (1997)
[astro-ph/9609169].

\bibitem{lmw}
D. Langlois, R. Maartens, and D. Wands, Phys. Lett. B{\bf 489},
259 (2000) [hep-th/0006007].

\bibitem{junc}
S. C. Davis, Phys. Rev. D{\bf 67}, 024030 (2003) [hep-th/0208205];
E. Gravanis and S. Willison, Phys. Lett. B{\bf 562}, 118 (2003)
[hep-th/0209076]; J. P. Gregory and A. Padilla, Class. Quantum
Grav. {\bf 20}, 4221 (2003) [hep-th/0304250].

\bibitem{cd1}
C. Charmousis and J-F. Dufaux, Class. Quantum Grav. {\bf 19}, 4671
(2002) [hep-th/0202107].

\bibitem{mt}
K. Maeda and T. Torii, Phys. Rev. D{\bf 69}, 024002 (2004)
[hep-th/0309152].

\bibitem{ln}
J. E. Lidsey and N. Nunes, Phys. Rev. D{\bf 67}, 103510 (2003)
[astro-ph/0303168].

\bibitem{footnote}
This bound results from viewing the GB contribution to the action
as a small correction, so that Eq.~(\ref{al}) holds, as one may
expect from a string-inspired perspective. In this case, the GB
term in the action would couple to the dilaton (in addition to
compactification moduli), and the bound would be consistent with a
small string coupling.

\bibitem{deser}
S.~Deser and A.~Waldron, Phys.\ Lett.\ B{\bf 508}, 347 (2001)
[hep-th/0103255].

\bibitem{fk}
A. Frolov and L. Kofman, hep-th/0209133.

\bibitem{tanmon}
T. Tanaka and X. Montes, Nucl. Phys. B{\bf 582}, 259 (2000)
[hep-th/0001092].

\bibitem{gens}
U. Gen and M. Sasaki, Prog. Theor. Phys. {\bf 105}, 591 (2001)
[gr-qc/0011078].

\bibitem{gs}
J. Garriga and M. Sasaki, Phys. Rev. D{\bf 62}, 043523 (2000)
[hep-th/9912118].

\bibitem{cdd}
C. Charmousis, S. C. Davis, and J-F. Dufaux, J. High Energy Phys.
{\bf 12}, 029 (2003) [hep-th/0309083].

\bibitem{cks}
F. Cooper, A. Khare, and U. Sukhatme, Phys. Rep. {\bf 251}, 267
(1995) [hep-th/9405029].

\bibitem{cho}
I.~Cho, hep-th/0402125.

\bibitem{cd}
C. Charmousis and J-F. Dufaux, hep-th/0311267.

\bibitem{fk2}
A. V. Frolov and L. Kofman, Phys. Rev. D{\bf 69}, 044021 (2004)
[hep-th/0309002].

\bibitem{grs}
D. S. Gorbunov, V. A. Rubakov, and S. M. Sibiryakov, J. High
Energy Phys. {\bf 10}, 015 (2001) [hep-th/0108017].

\bibitem{kkt}
T.~Kobayashi, H.~Kudoh, and T.~Tanaka, Phys.\ Rev.\ D{\bf 68},
044025 (2003) [gr-qc/0305006].

\bibitem{llkcba}
J. E. Lidsey, A. R. Liddle, E. W. Kolb, E. J. Copeland, T.
Barreiro, and M. Abney, Rev.  Mod. Phys. {\bf 69}, 373 (1997).

\bibitem{tg}
S. Tsujikawa and B. Gumjudpai, astro-ph/0402185.

\bibitem{hl1}
G. Huey and J. E. Lidsey, Phys. Lett. B{\bf 514}, 217 (2001)
[astro-ph/0104006].

\bibitem{hl2}
G. Huey and J. E. Lidsey, Phys. Rev. D{\bf 66}, 043514 (2002)
[astro-ph/0205236].

\bibitem{gklr}
G. F. Giudice, E. W. Kolb, J. Lesgourgues, and A. Riotto, Phys.
Rev. D{\bf 66}, 083512 (2002) [hep-ph/0207145].

\bibitem{st}
D. Seery and A. N. Taylor, astro-ph/0309152; G. Calcagni, JCAP
{\bf 11}, 009 (2003) [hep-ph/0310304].

\bibitem{liddle}
S. M. Leach and A. R. Liddle, Phys. Rev. D{\bf 68}, 123508 (2003)
[astro-ph/0306305]; S. Tsujikawa and A. R. Liddle, JCAP {\bf 03},
001 (2004) [astro-ph/0312162].

\bibitem{mwbh}
R. Maartens, D. Wands, B. A. Bassett, and I. P. C. Heard, Phys.
Rev. D{\bf 62}, 041301 (2000) [hep-ph/9912464].

\bibitem{wmll}
D. Wands, K. A. Malik, D. H. Lyth, and A. R. Liddle, Phys. Rev.
D{\bf 62}, 043527 (2000) [astro-ph/0003278].

\bibitem{tsm}
S. Tsujikawa, M. Sami, and R. Maartens, astro-ph/0406078.


\end{thebibliography}
\end{document}